%% file: main.tex
\documentclass[10pt,twocolumn, smallerheadings]{article} 

\usepackage{cite}
\usepackage{amsmath,amssymb,amsfonts}
\usepackage{amsmath,amssymb}
\usepackage{algorithmic}
\usepackage{graphicx}
\usepackage{textcomp}
\usepackage{xcolor}
\usepackage{booktabs}

\input{preamble}

\anonymousfalse
\longtrue
\commentstrue 

\title{The Vulnerable Nature of Decentralized Governance in DeFi}
\author{Maya Dotan, Aviv Yaish, Hsin-Chu Yin, Eytan Tsytkin, Aviv Zohar\\ Hebrew University of Jerusalem, Israel}

\begin{document}

\maketitle

\begin{abstract}
    Decentralized Finance (DeFi) platforms are often governed by Decentralized Autonomous Organizations (DAOs) which are implemented via governance protocols. Governance tokens are distributed to users of the platform, granting them voting rights in the platform's governance protocol.
    Many DeFi platforms have already been subject to attacks resulting in the loss of millions of dollars in user funds.
    
    In this paper we show that governance tokens are often not used as intended and may be harmful to the security of DeFi platforms.
    We show that (1) users often do not use governance tokens to vote, (2) that voting rates are negatively correlated to gas prices, (3) voting is very centralized.
    
    We explore vulnerabilities in the design of DeFi platform's governance protocols and analyze different governance attacks, focusing on the transferable nature of voting rights via governance tokens. Following the movement and holdings of governance tokens, we show they are often used to perform a single action and then sold off. We present evidence of DeFi platforms using other platforms' governance protocols to promote their own agenda at the expense of the host platform. 
\end{abstract}


\section{Introduction}\label{section:intro}

Cryptocurrencies such as Bitcoin~\cite{nakamoto2008bitcoin} and Ethereum~\cite{wood2014ethereum,buterin2022ethereum} facilitate monetary \emph{transactions} between users in a distributed and decentralized manner. Users who wish to have their transaction processed by the system can broadcast it to entities called \emph{miners}, who in turn collect transactions in \emph{blocks}. As transactions are ordered within each block, and as blocks contain a reference to at least one preceding block, an ordered ledger of transactions commonly called a \emph{blockchain} is formed.
Various mechanisms such as Proof of Work (PoW)~\cite{dotan2021proofs} and Proof of Stake (PoS)~\cite{bano2019sok} are used to maintain the integrity and security of the ledger. The differences between them are outside the scope of our work, but we note that exact terminology used by each might differ, too. For example, in Ethereum, miners are also called \emph{validators}. For brevity, we will stick to the terminology as used in Bitcoin.

The security of blockchain protocols relies on blocks being quickly propagated throughout the network \cite{dotan2021survey}, therefore requiring a size-limit on blocks which limits the system's throughput. Thus, if the amount of pending transactions exceeds the maximal throughput, users can prioritize a transaction over competing ones by offering a \emph{fee} to the first miner to include it in a block~\cite{gafni2022greedy}. 

The Ethereum blockchain allows transactions to contain software programs specified using a formal virtual-machine~\cite{goldberg1974survey} execution model called the \emph{EVM} (Ethereum virtual machine)~\cite{wood2014ethereum}. These programs, also known as \emph{smart contracts}~\cite{antonopoulos2018mastering}, can be \emph{deployed} to the blockchain, e.g., stored on it, thereby allowing users to interact with them by creating transactions that invoke their functions. Transaction fees on the Ethereum network are often referred to as ``gas''. At times of high demand, gas prices for including transactions are high, and at times of low demand they drop. 

\subsection{Decentralized Finance}
\label{section:decentralizedFinance}
DeFi platforms, that implement traditional financial instruments on top of a decentralized mechanism, have emerged as a leading use-case for smart contracts on the Ethereum blockchain~\cite{wood2014ethereum,buterin2022ethereum}. In this work, we will focus on two types of DeFi platforms: \emph{DEX}s (Decentralized exchanges) and decentralized lending platforms.

DEXs such as Uniswap~\cite{uniswap2022uniswap} enable users to exchange or \emph{swap} tokens amongst each other without requiring any form of direct interaction between the exchanging parties~\cite{xu2021sok}. These tokens are commonly implemented using the ERC-20 standard~\cite{moin2020sok,victor2022measuring}, which is an inter-operable specification for fungible tokens~\cite{ethereum.org2021erc,vogelsteller2015eip}.

Decentralized lending platforms such as Aave~\cite{wow2020aave} and Compound~\cite{leshner2019compound} let users take and give loans.
There are two main types of loans available on DeFi platforms (1) long-term \emph{collateralized} loans 
which are secured by up-front deposits (e.g., collateral)~\cite{yaish2022manipulating},
and (2) \emph{flashloans}~\cite{daian2020flash} which are loans given for the duration of a single transaction.
The transaction atomicity offered by the EVM allows one to ensure that if a flashloan is not repaid by the conclusion of the transaction that took it, the transaction is reverted~\cite{beregszaszi2017eip}.

In this paper we chose to focus on the Ethereum blockchain, as it is do date the largest and most active blockchain in terms of DeFi TVL. We also focus on the following DeFi platforms: Aave V2, Uniswap, Compound, Balancer. At the time of writing this paper, these are the biggest DeFi platforms with an active governance protocol, with a combined total value locked (TVL) of over $12$B USD. They are also the platforms with the highest value governance tokens on the market at the time of writing this paper.

\subsection{Governance Protocols} \label{section:governanceProtocols}
\begin{figure}
    \centering
    \includegraphics[width=0.8\columnwidth]{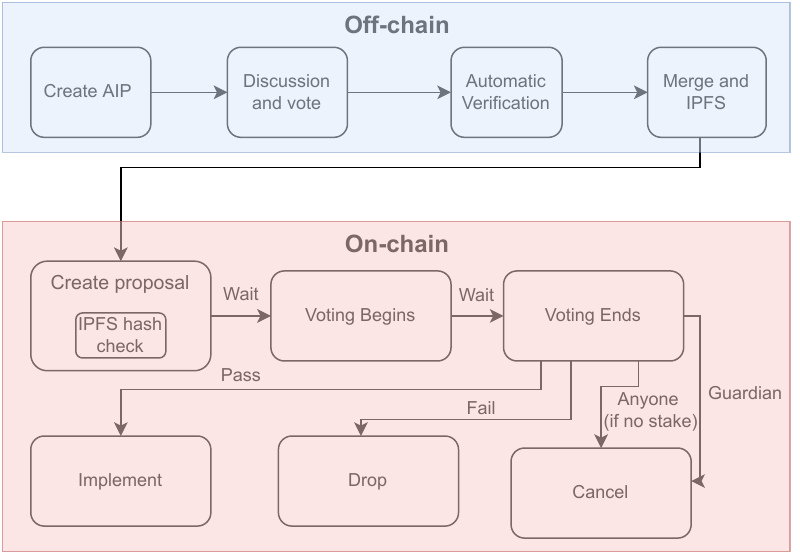}
    \caption{Aave's governance proposal pipeline. There are both on and off chain components to the process. At the point of moving from the off-chain to the on-chain process, the proposer must give proof of passing the off-chain process in the form of an IPFS hash. Uploading to IPFS can only be done by Aave's team.}
    \label{fig:AaveGovernancePipeline}
\end{figure}

The management of funds in DeFi platforms is often done via DAOs. This is done algorithmically via smart contracts that are deployed on the blockchain. Smart contracts that follow the \emph{proxy design pattern}~\cite{marchesi2020design}
appoint admins that can change the address of the delegate contracts. These admins need not be humans. For example, the admins for the various of Compound's smart contracts are smart contracts themselves which enable decentralized decision-making \cite{leshner2019compound,finance2022governance}. Such contracts are also called \emph{governance protocols}~\cite{jensen2021decentralized,gudgeon2020decentralized}.  These contracts are publicly visible and audit-able to anyone who has the technical skill to read them. There are typically some aspects of platform management that remains alterable over time, to adjust the platforms to market events, infrastructure upgrades and so on. 
While these platforms are typically privately owned and managed, they many times delegate many aspects of the managements of funds and decisions regarding crucial policy changes to their users. This is done in an attempt to increase the perceived transparency and decentralization of these platforms. This is what is typically referred as a \emph{governance protocol}. 
Governance protocols of different platforms vary. Some governance protocols involve both on and off chain steps. The execution and implementation is done on-chain. In~\cref{fig:AaveGovernancePipeline} we demonstrate the steps involved in life-cycle of a proposal in Aave's governance protocol. Off chain activity mainly consists of introducing pending change proposals to the community and open discussion about their implications. 

Practically, this delegation of decision making is often done through the utilization of \emph{governance tokens}. These tokens dispersed to community members as a reward for active participation in the platform, through depositing or lending funds, making exchanges etc.~\cite{fritsch2022analyzing}. The tokens then become trad-able on the same markets that they enable controlling. 
Decentralized governance in the on-chain setting broadly implies one token equals one vote. Governance tokens can be earned, and can be printed by the platform. Voting is done via an on-chain transaction, and thus incurs a cost due to gas fees. Votes have monetary value, people might not vote because their tokens are put to other uses, e.g. as collateral, or even selling your vote to make a monetary profit.
Votes are often on topics which don't directly affect the voter, e.g. change the parameters for some token which the voter doesn't hold, and information regarding the content and implications of the votes may be vague. In fact, many times a voting party may introduce a new policy change on which other parties can vote, and make the description intentionally misleading. It is left to other voters to individually understand the implications of the suggestion. Votes can be automated using bots and can be delegated to others (either explicitly, or by loaning out your tokens).

Note that governance token can be traded on the free market, just as any other token. This means that individuals can buy voting shares of the platforms even without actively participating in the platform. 

\paragraph{Escrowed Governance Tokens}
\emph{Vote Escrowed Tokenomics} is a different way of issuing governance power to users, meant to increase the commitment of governance voter and proposers in the longevity and health of the platform.
In the vote escrow mechanism, users lock tokens in the platform, and in return are rewarded \emph{veTokens}, representing voting rights. Tokens are distributed according to the time a user committed to locking the funds on the platform, longer locking periods grant higher voting power. The purpose is to align users' incentives with the health of the platform. Several DeFi platforms now use veTokens in their governance protocol, including Curve.fi \cite{curve2022veTokens} and Balancer \cite{balancer2022veTokens}. 

\subsection{Our Contributions}\label{sec:our_contribution}
In this paper we show misuse of governance protocols by bad actors. In particular, changes made to the platform are not always at the benefit of the platform, and are sometimes made by user who only hold on to their voting rights for the duration of a single proposal.
We find that these protocols often do not encourage active participation in the governance mechanisms based on real-world data about voting patterns which indicate low participation rates and that voting is very homogeneous. This enables bad actors to gain disproportional voting power in these protocols. 
We map out governance attacks in which individuals use governance protocols to attack the platform they control. We show that the vested escrow governance design is not immune to this type of abuse.
Finally, we identify platforms that utilize other platform's governance mechanisms in order to promote their own interest, sometimes at the expense of the native platform. 


\paragraph*{Paper structure}
In~\cref{section:voting} we analyze voting patterns in popular DeFi platforms. In~\cref{sec:vulnerability} we explore the vulnerability of governance attack to misuse: In~\cref{sec:governance_attacks} we detail some recent attacks on DeFi platforms that utilize the platform's own governance mechanism to the attacker's advantage. In~\cref{sec:cross platform} we examine cross-platform governance activity. We focus both on platforms holding significant stake in other platforms, and even proposing changes to other platform to promote their own agenda. In~\cref{section:relatedWork} we review related work, and conclude in~\cref{section:conclusions}.




\section{Governance Voting and Proposing Patterns}
\label{section:voting}
In this section, we dive deeper into voting patterns in several main governance protocols. We analyze the holding and voting patterns of users and find evidence of voting centrality in the sense that users typically do not vote against proposals. This can be explained by the fact that voting on proposals costs funds reflected in transaction fees. Fees vary according to network congestion. An attacker can leverage this to propose malicious attacks at times of high congestion, making the attack prevention more expensive for users.

In~\cref{table:voingResults} we see an overview of several key aspects of the four major DeFi platforms examined in this paper. 
Balancer is the only platform to utilize escrowed governance tokens, making the voting rights not trade-able for a period of time.
We see that all four platforms have a point of centralization in the Guardian entity. This entity, controlled by a \emph{community multisig}, has the power to cancel proposals. The multisig is composed of several users that are believed by the creators of the platform to hold the platform's best interest at heart, and to cancel malicious proposals. For instance, in Aave this multisig requires 6 out of 10 of the holders to sign an order to kill an active proposal. This is a point of centralization which can be a safeguard, but also might be subject to abuse\footnote{For instance, the famous Ronin bridge hack involved corrupting 5 of the 9 required shares~\cite{lossless2020roninmultisig}}. Additional points of centralization are found in integrating a platform's on/off chain proposal process. For instance, In Aave, once a proposal passes the off-chain stage it needs to be merged into Aave's repository in order to generate a valid IPFS hash for uploading to the blockchain, see~\cref{fig:AaveGovernancePipeline}. This upload can only be done by Aave's developers. Except Balancer, all platforms have some measures of automatic code validation for pending proposals~\cite{bgd2020seatbelt}. All platforms also have some form of execution delay between the time a proposal is uploaded, is voted on and is then implemented to prevent flash-loan attacks, which we discuss in~\cref{sec:governance_attacks}.
Both of these safe-rails are meant to defend the platform from malicious attacks. However, they are not bulletproof. For instance, later in this section we detail cases of users who only purchase and hold on to a large portion of governance tokens for the duration of a single proposition of their own making. They vote to pass these proposals and then transfer on their governance tokens. In~\cref{sec:cross platform} we dive deeper into these use-cases and show that these were proposals that were harmful to the platform. These were not stopped by the safe-rails, and they past with overwhelming majority, one of them even past unanimously.

\begin{table*}[h]
	\centering
	\caption{An overview of popular governance platforms, centralization points, proposal process flow and centrality in voting.}
	\label{table:voingResults}
	\scalebox{0.8} {
	\begin{tabular}{c c c c c c c}
		\toprule
		
		\thead{Platform} & 
		\thead{Proposals can be \\ cancelled by Guardian} &
		\thead{Combines On and \\ Off-chain Protocols} &
        \thead{Automatic Proposal \\ Code Validation} &
        \thead{\# Proposals (\% Passed with \\99\% majority)} \\
		
		\midrule
		
		Aave &
		$\checkmark$ &
		$\checkmark$ &
        $\checkmark$ &
		$109$ ($76.1\%$) \\
		
		Compound &
		$\checkmark$ &
		- &
		$\checkmark$ &
		$134$ ($61.7\%$) \\
		
	    Uniswap V3 &  
		$\checkmark$ &
		$\checkmark$ &
        $\checkmark$ &
		$24$ ($66.7\%$) \\

        Balancer &  
		$\checkmark$ &
		$\checkmark$ &
            - &
		$286$ ($61.2\%$) \\

		
		\bottomrule
  
	\end{tabular}
	}
\end{table*}

\begin{table}[]
\center
\caption{Percentage of circulating governance tokens used for voting, by platform.
Evidently, Compound users feel more incentivised to vote than Aave's. Also, Uniswaps' usage of governance proposals is significantly lower the the other two platforms.}
\label{PctOfCirculationUsedForVoting}
\scalebox{0.8}{
\begin{tabular}{lccc}
\toprule
\textbf{} & \textbf{Aave} & \textbf{Compound} & \multicolumn{1}{l}{\textbf{Uniswap}} \\
\midrule
Proposals & 109           & 134              & 24                                    \\
H-Index   & 86            & 91               & 16                                    \\
\midrule
Mean      & 3.67\%        & 11.21\%          & 6.41\%                                \\
Std. dev  & 2.92\%        & 5.64\%           & 3.81\%                                \\
Min       & 0\%           & 0\%              & 0.00\%                                \\
Median    & 3.15\%        & 10.62\%          & 6.64\%                                \\
Max       & 23.53\%       & 25.97\%          & 11.93\%                               \\
\bottomrule
\end{tabular}}
\end{table}

\subsection{Voting and proposal centrality in Aave, Compound and Uniswap} 
\cref{PctOfCirculationUsedForVoting} summarizes proposition and voting data from three major platforms. In all three platforms the percentage of governance tokens that participate in voting is very low, with Compound leading the chart at $11.2\%$.  In~\cref{PctOfCirculationUsedForVoting} we also see that 
an overwhelming majority of proposals pass with an overwhelming majority of votes in favor. We look at the equivalent of an ``h- index'' of the three platforms, i.e. calculated by counting the percent of proposals that won at least that same percent of votes. In Aave, for instance, at least 86 in the sense that there are 86\% of the proposals with at least 86\% of the votes.
In~\cref{table:voingResults} we also see indication of voting centrality, in all platforms a majority of all proposals passes with over $99\%$ majority (out of the total tokens used to vote, not of all available tokens). Combining these results we have that voting participation is low and the user that do vote, do so in a homogeneous manner.



\begin{figure}[h!]
    \centering
	\includegraphics[width=0.9\columnwidth]{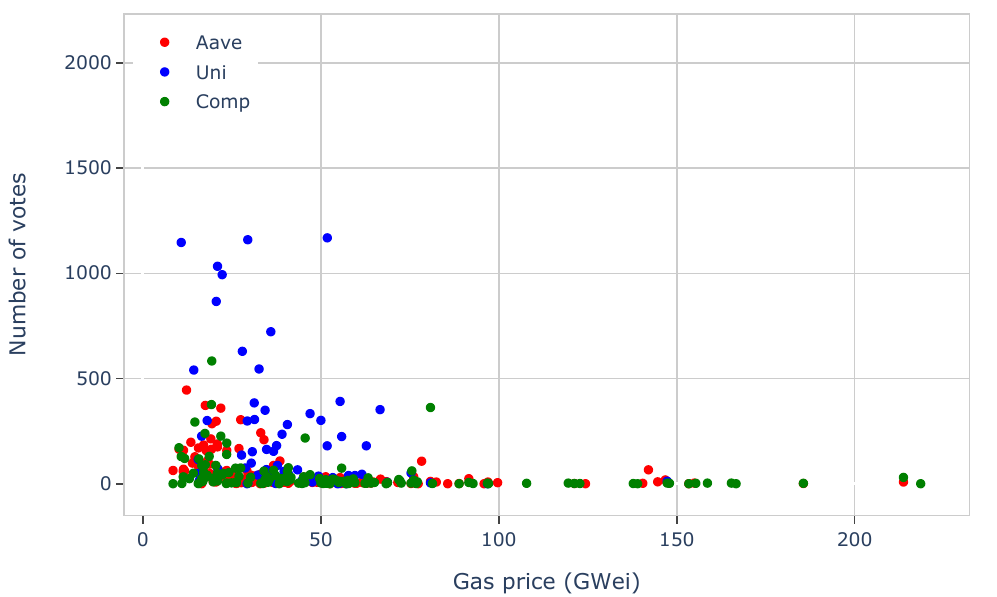} 
    \caption{Number of votes cast during times of active governance proposals vs. the Ethereum gas price in Aave, Compound and Uniswap. In all three platforms the gas price is negatively correlated with the number of votes cast on governance proposals.}
    \label{fig:votingVsGasPrice}
\end{figure}

A plausible explanation is that casting a vote in a governance protocol requires creating an on-chain transaction, which incurs transaction fees. These fees vary according to network congestion. This means that the price of voting changes with blockchain conditions that be unrelated to the DeFi platforms and certainly to the governance protocols. In~\cref{fig:votingVsGasPrice} we see indication that there is negative correlation between the Ethereum gas price and the number of votes cast for active proposals, which could imply that voters tend to think twice about voting when the price of voting is high. This also introduces as new avenue for attacker to upload governance proposals at times of high transaction fees.
\cref{table:voingResults} also summarizes additional points of centralization in the proposal process in the different platforms.

\subsection{Governance Token Holding Patterns and Movement in Aave}
\begin{figure*}[h!]
    \centering
    \subfloat[Proposers in Aave - This image depicts AAVE token holdings of governance proposers in Aave over time. The crossed lines indicate if a holder used their tokens to propose a governance change at the given time. Shown are only users that have ever used their tokens to propose. We see that proposals 27 and 65 we made by users that only held on to these tokens for the duration of a single proposals life-cycle (proposing and voting).]{
	\label{fig:holders_proposers}
	\includegraphics[width=0.9\columnwidth]{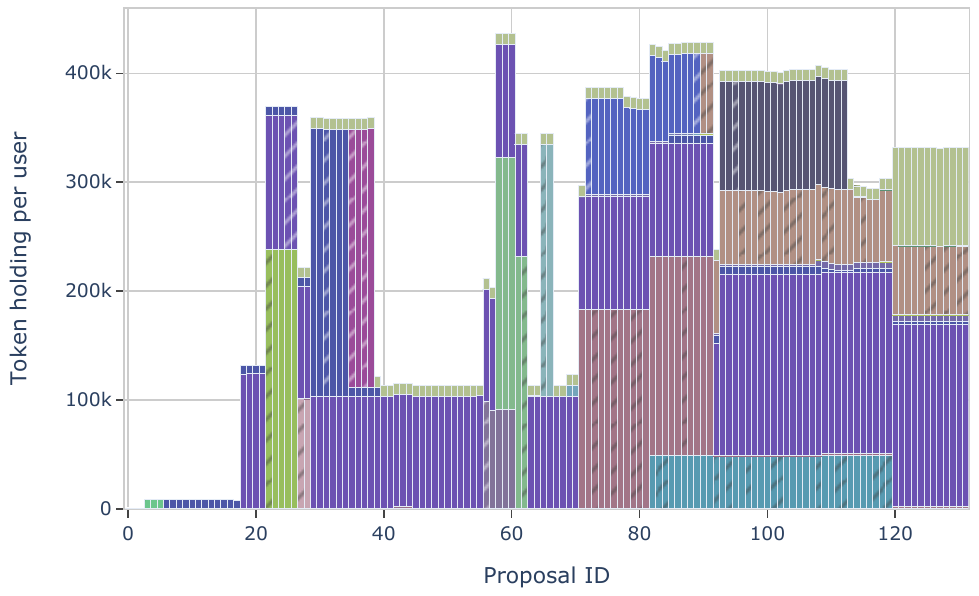} } 
 \qquad
    \subfloat[Voters in Aave - This image shows the top 600 AAVE token holders over time and their voting activity. Excluded from this image are Aave's own reserves pool and the Aave genesis team (whom are the majority stakeholders in the platform). The crossed lines indicate if a holder used their tokens to vote a governance change at the given time. We see participation rates in voting vary, and users choose on which proposals they care to vote. None of the users vote on all proposals that happened during the time they held on to their AAVE tokens. We also see that out of the total $14$M available AAVE tokens, the number of tokens used to vote is very low, typically under $500$K tokens]{
	\label{fig:holders_voters}
	\includegraphics[width=0.9\columnwidth]{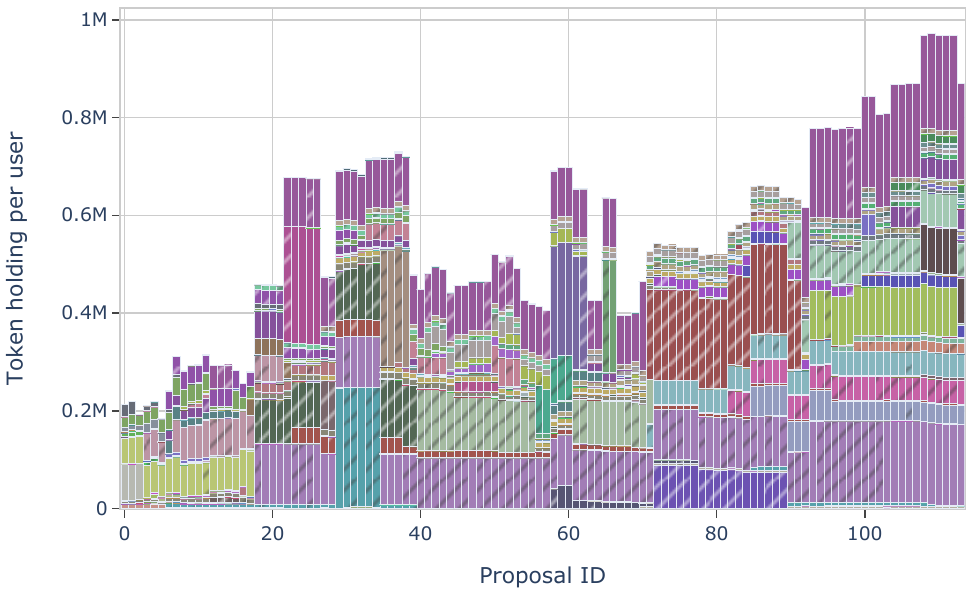} } 
\caption{AAVE token holdings over time. We watch how these tokens are transferred between holders at times proposals for proposers(left) and voters (right).}
    \label{fig:aave_holders}
\end{figure*}

As a case study, we collected data on governance token holders in Aave. We looked at how these holdings change over time, and whether the token holders indeed participate in the governance process, both in proposing and voting. Results are summarised in~\cref{fig:holders_voters}, \cref{fig:holders_proposers}. Excluded from \cref{fig:holders_voters} are (1) Aave's reserve pool, the largest token holder out of the top 600 holders. Over the entire life of the protocol, the reserve tokens were only used once to vote in proposal 106\footnote{``Adjust Aave Governance Level 2 requirements'' which changed the structure of the governance protocol itself (lowering the quorum requirements)}, and was dormant for all of the rest. 
(2) Aave's genesis team.
In~\cref{fig:holders_proposers} we see that that there are several cases of users who only held on to governance tokens for the duration of the proposal they introduces, and transferred them elsewhere upon completing their proposing business. This implies that the governance protocol might not always be utilized by long term stake holders of the platform, but rather by persons of interest that buy the tokens to promote a single proposal of interest to them and then dump them when the proposal ends its life-cycle.
We map these instances to several proposals that were passed and were eventually reversed by the community since they were harmful to the platform. This is detailed in~\cref{sec:cross platform}.

\begin{figure}[h!]
    \centering
	\includegraphics[width=0.9\columnwidth]{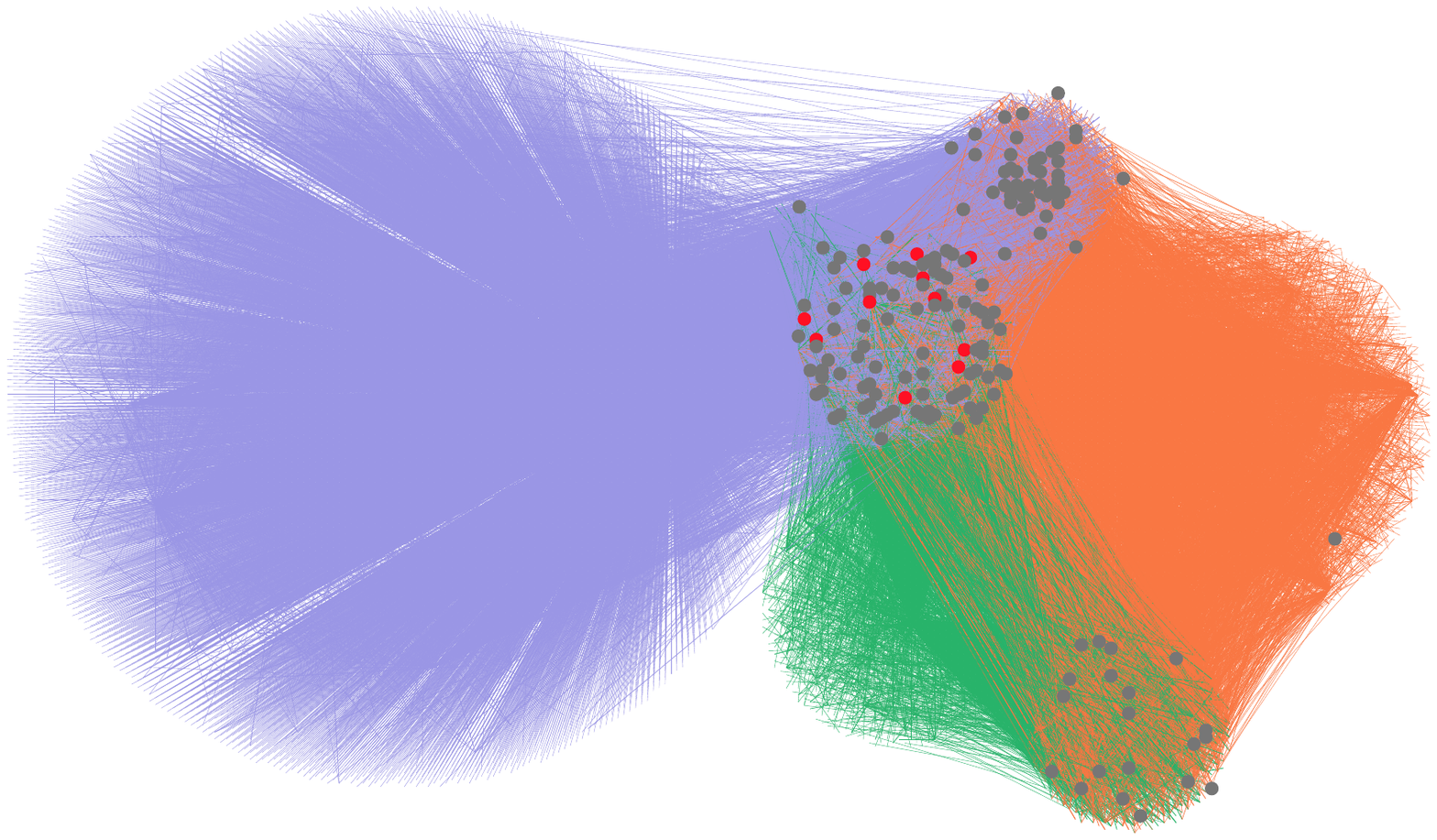} 
    \caption{Governance token movement for Aave (Purple), Compound (Green) and Uniswap (orange). The red nodes are voters on all three platforms, grey nodes voted in two out of the three.}
    \label{fig:tokens_movement}
\end{figure}

Data shows tokens move frequently trough DEXs and other DeFi platforms, explained by the fact that AAVE tokens earn interest in various DeFi platforms. Combined with evidence showing that voters and proposers only hold on to AAVE tokens for short periods of times when they are actively voting/proposing paints a picture of these tokens being a quick way for parties of interest to make vast changes to the code of the platform without necessarily having a long term interest in the health of the platform or even an interest in honestly participating in the platform for its intended use. Additionally, we identify users that vote in multiple platforms. 
In~\cref{fig:tokens_movement} We can see the movement of governance from the three platforms - Aave, Compound and Uniswap. In red are voters on all three platforms. They consist of two platforms that aid in voting/delegating votes (Sybil and Excelsior), the group ``Blockchains at UCLA'',  DeFi Pulse Index (whom we discuss in depth in~\cref{sec:cross platform}), and the rest are either pseudo-names or unidentified.

\section{Vulnerability of Governance Protocols}\label{sec:vulnerability}
In this section, we review use-cases of governance protocols being vulnerable to misuse by governance token holders. In~\cref{sec:governance_attacks} are outright attacks, scenarios in which the user manages to either steal funds from the platform by momentarily holding governance tokens, or manipulate the reward scheme. Use-cases in~\cref{sec:cross platform} are more subtle, the misuse is not obvious at first, and is hence not stopped by safeguards of the protocol. In retrospect they are sometimes harmful to the host platform, and require action by the community to remedy the situation.

\subsection{Governance Attacks and Risks}\label{sec:governance_attacks}
Any user with sufficient stake of the governance token can propose changes, in code, overriding the current smart contract implementation of the platform. Proposals are sometimes difficult to verify, especially for end users that are not proficient in reading code. In smart contracts, there are clever ways of disguising malicious code. These ways have been exploited by attackers in order to upload malicious proposals that are seemingly harmless. Governance attacks are a growing way for bad actors to steal funds from DeFi platforms. These attacks are often very well hidden and incredibly sophisticated, and are not discovered by the various safety measures implemented by platforms to prevent them.

\subsubsection{Balancer and Humpy}
Balancer is a DeFi platform that uses vote-escrowed governance via veBAL tokens. veBAL tokens are distributed to users who commit to locking funds into Balancers' platform. The ``voting strength'' a user has is determined both by the quantity of the locked tokens and the duration funds are locked. The voting strength decays on a weekly basis (as the funds approach being unlocked). 
This design means that committing to holding BAL long term will give an advantage to in strength to long term holders, and is supposed to increase the stake of governance voters hold in the health of the platform over time.
veBAL holders can obtain governance rights, and accrue a portion of protocol revenue.

Following the introduction of veBAL to Balancers' governance, a user under the pseudoname \emph{Humpy} managed to get control of 35\% of the entire supply of veBAL tokens. 
Humpy leveraged veBAL to create a CREAM/WETH pool and set the pool trading fees to 10\%. Then, Humpy directed its veBAL holdings across multiple addresses and directed BAL emissions to the pool’s gauge. This way, Humpy managed to direct 1.8M USD of cumulative BAL emissions to itself, via the to the CREAM/WETH gauge. This was a great profit for Humpy, at the expense of Balancer's platform, as they only generated ~17K USD in revenue from the new pool over the same period.

\begin{figure}[h!]
    \centering
	\includegraphics[width=0.9\columnwidth]{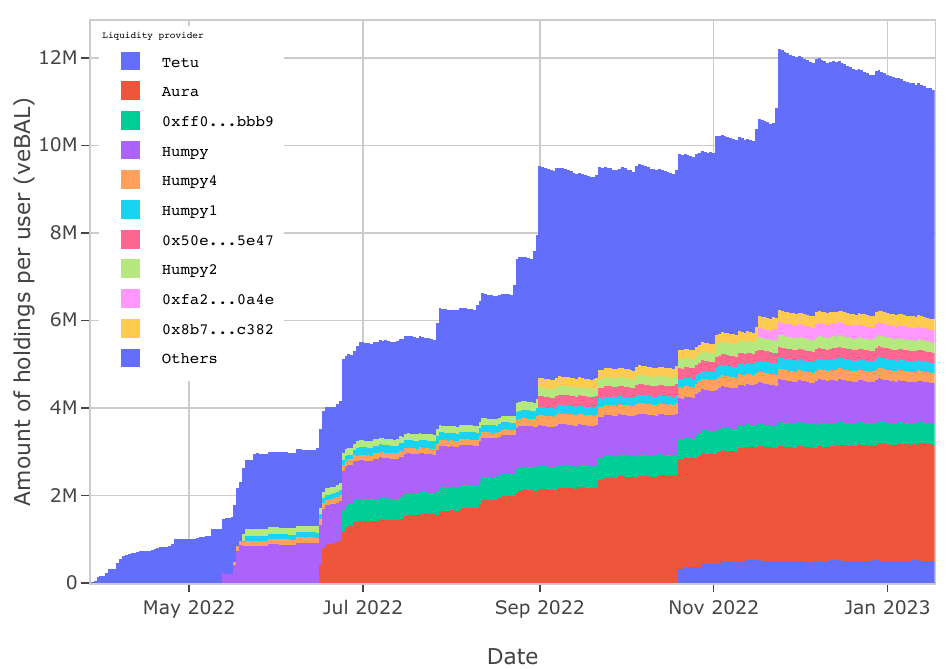} 
    \caption{veBAL token holders over time. Each color represents a specific user. There are several addresses associated with Humpy. Attribution of addresses to Humpy was done by looking at voters that voted against proposal BIP 28 (Kill CREAM/WETH Gauge), a proposal which stopped Humpy's original money printing pool. Humpy is also accountable for the growing portion of the Tetu pool.}
    \label{fig:veBAL_holders}
\end{figure}

The Balancer community noticed this occurrence mitigated the problem by issuing proposal \href{https://messari.io/governor/proposal/30f64464-af87-47a9-8e8a-6a12b4cf12b8}{BIP 19} at an attempt to better align the veBAL design with Balancer's revenue. 
The proposal passed and Balancer introduced several new safeguards.
Following this, Humpy changed strategies, moving away from low-cap pools into the Tetu 20WETH/80BAL/TetuBAL pool. Which again caused Humpy to be a fast printer of BAL tokens.
This created a proposition war between Humpy and the Balancer community. 
The saga came to an end with a peace treaty proposal~\cite{messari2022peaceTreaty} between Balancer and Humpy, where 
both parties agreed to give Humpy control of 17.5\% of all future BAL emissions.
We visually present the data of the distribution of Balancers veBAL token holders over time in~\cref{fig:veBAL_holders}. We can see how Humpy's stake, as well as increasing amounts of tokens being transferred to Tetu, which is also associated with Humpy.

In this case, an oversight in the design of the governance mechanism created an opportunity for a selfish, utility maximizing user to leverage the Balancer's governance to grow its own wealth at the direct expense of the wealth and stability of the platform. This was only resolved by surrendering significant power permanently to the attacker, essentially making it a major stakeholder indefinitely. Clearly, this is an undesirable outcome from a mechanism that is supposed to manage the health of the platform.

\subsubsection{Beanstalk}
Beanstalk is a lending platform that allowed its governance contract to change the address of the owner of the collateral of the platform. It was attacked \cite{cointelegraph2021beanstalk} by an exploiter 
who transferred all of the collateral of the platform (estimated at just over 182M USD) to themselves, collapsing the platform as a result. The attack was issued by an individual who was able to take a flashloan of Beanstalk's governance tokens and in doing so became a majority holder of governance tokens, allowing them to make dictatorship decisions in the governance contract. The attacker was able to upload a seemingly innocent governance suggestion (donating funds to Ukraine) while hiding the actual functionality using Ethereum's Diamond standard \cite{mudge2022eip} (which is a sophisticated version of the proxy mechanism) and was also able to use the emegranctCommit() method to transfer to himself a sum that covered his flashloan, alongside over 180M USD in profits from the attack. 

\subsubsection{Tornado Cash}
Tornado Cash~\cite{pertsev2019tornado} is a popular Ethereum based zero-knowledge mixer implemented as a smart contract. The platform is governed by a DAO via the TORN governance token. On May 20th 2023 the DAO handling operations, funds and future plans of Tornado Cash was taken over by an unidentified attacker~\cite{coindesk2020TORNattack}. 
The attacker uploaded a malicious proposal that hid a code function that granted them fake votes. The attackers proposal imitated an earlier version – except with some malicious code that allowed for the update of logic that gave the attacker access to all governance votes.

\subsubsection{Compound COMP and negative interest rates on loans}
In two distinct incidents, Compounds' governance reward mechanism (i.e. COMP token distribution) created condition in which the interest rate on COMP loans in Compounds' own platform were negative. In the first case, COMP rewards rates were applied at the same rate for both suppliers and borrowers for any single market, which created negative interest rates when borrowing certain assets from the platform. Governance proposal 62 \cite{compound2021proposal62} changed the distribution to fix the problem.
Proposal 68 \cite{compound2021proposal68} was imposed to remedy another negative interest situation. 
Compound used to reward cCOMP\footnote{this token is printed by Compound to represent positions in the COMP token} borrowing with COMP tokens. This meant the net rate for borrowing COMP was negative. 
In both these examples, a user that borrows COMP token both makes revenue on \emph{borrowing} funds and increases their voting power in the platform, as the COMP token is the governance token used to vote on and propose changes to the Compound protocol.
In addition to being bad for the platform's balance, the fact that this happened with the platform's governance token made the risk worse, since the negative interest rate de-facto meant that users borrowing COMP tokens disproportionately increased their voting power in the platform.

\subsection{Cross-Platform Governance activity}\label{sec:cross platform}
In this section we present evidence of platforms holding stake in other platforms' governance. This stake can either be bought by a competing platform, or sometimes be gained by simply holding user's tokens which happen to be governance tokens in other platforms. This creates dependencies between platforms, exposing them to devastating risks.
We present two examples of proposals in Aave's governance protocol in which users leveraged Aave's governance mechanism to propose and vote on a single transaction to promote their own token on Aave's platform. Both proposals passed and were implemented without much objection. Both were later overturned upon new government proposals, due to  extreme drops in the value of the tokens, that were substantial enough to risk the health of the Aave ecosystem.

\subsubsection{Terra's UST in Aave}
Terra is a blockchain protocol and payment platform for algorithmic stable-coins. 
On March 8th 2022 proposal number $65$ was created on Aave's governance protocol~\cite{aave2022proposal65}. The title of the proposal is ``Add Terra USD (UST) to Aave v2''. The proposer of this transaction is Ethereum address \href{https://etherscan.io/address/0xf1a39ccaa7b3e0a6f82c0806518925b54024f011}{$0xf...011$}. This also the biggest voter for the transaction, with $230.9$K out of the total $487.7$K votes in support of this proposal. The proposal passed and was executed on Aave's platform. The proposer got ownership of the AAVE tokens just before this proposal, and discarded them right after the vote. The Terra USD token collapsed on May 12th 2022 after it depegged from the USD, an event that wiped out almost \$45 billion in market capitalisation within a week\cite{bloomberg2022terraluna}. Adding the UST token to Aave's platform  did not promote the health of the system. During the process of preparing the proposal on-chain this address received funds from address \href{https://etherscan.io/address/0xbec5e1ad5422e52821735b59b39dc03810aae682}{$0xbe...e682$} which is associated with Terraform Labs, the company behind the now collapsed Terra stable-coin UST, hinting that Terraform Labs was behind this proposal. This in potentially catastrophic to Aave's platform, as loans taken out with UST as collateral will most likely never be repaid, as the collateral is now worthless. 

Aave only noticed this after the collapse.
On May 19th 2022, the Aave DAO executed proposal 75~\cite{aave2022proposal75}, this time for ``Freezing UST and Updating stETH Parameters''. This proposal also passed with $450$K votes in favor of freezing the asset on Aave's platform.

\subsubsection{DeFi Pulse Index and Meta-Governance}
The DeFi Pulse Index (DPI) is a DeFi platform aiming to implement a decentralized capitalization-weighted index that tracks the performance of DeFi assets on the Ethereum blockchain. DeFi Pulse index purchased different ERC-20 tokens across the Ethereum DeFi ecosystem, 
in the native tokens to the relevant platforms. In Aave, Compound, Balancer and Uniswap the native token is also a governance token. DPI is aware of this and has implemented a ``Meta-governance'' protocol. Currently, this enables DPI holders to vote on changes to the Aave, Compound, and Uniswap (DPI stated they will expand to other protocols in the future). Interestingly, the only governance proposal that DPI holder have voted on in the Aave platform was proposal 27~\cite{aave2021proposal27} to ``begin the on-boarding process for listing DeFi Pulse Index (DPI) as collateral on the Aave ARC market''. 
The proposal was created by Ethereum address \href{https://etherscan.io/address/0x7f4C5938AF9731e9feadc09C3FA782508198532E}{$0x7f\ldots 32E$} , an address funded by DPI in transaction \href{https://etherscan.io/address/0x37e6365d4f6ae378467b0e24c9065ce5f06d70bf}{$0x37\ldots 0bf$}. This proposal whitelisted DPI on the Aave platform, making it viable as collateral to loan-takers on the Aave platform. 
%
Later, in proposal $189$~\cite{aave2023proposal189} ``Freeze DPI on V2 Ethereum'' Aave decided to stop support for DPI due to ``implied centralization risk of DPI compared with the direct holding of the basket of underlying assets''. 
DeFi Pulse Index was able to gain enough AAVE governance tokens to momentarily vote their own token, DPI, into Aave's protocol. It took the Aave community over a year to realize that having DPI as collateral in Aave is harmful to the platform and the ecosystem due to increased risk of cascade in case of a drop in DPI's value, and correct it.

\subsubsection{Platforms Holding Stake in Other Platforms}
In some cases, platforms choose to hold stake in other platforms in what is meant to be a benign partnership. The nature of governance tokens however, means these stake-holdings can cause harm to the health of at least one of the platforms. Two notable examples are:

\paragraph{Binance and Uniswap}
On Oct. 2022, Binance~\cite{binance2021binance} shifted $~4.6M$ UNI tokens between two different wallets owned by Binance. This caused $13M$ UNI tokens to be automatically delegated to Binance's wallet on Uniswap by accident\cite{coindesk2022binanceUni}. This made Binance one of the largest delegates in Uniswap at the time. Although Binance did not vote with the UNI tokens, they could have. This is an example of how a bug in Binance's platform directly affects Uniswap's governance.

\paragraph{Aave and Balancer}
The Aave community has passed two governance decisions (proposal 87~\cite{aave2022governanceProposal87} and proposal 115~\cite{aave2022governanceProposal115} to buy stake in BAL tokens, as part of a strategic partnership between the platforms. They are aiming to re-lock these tokens back into Balancer in order to gain stake in veBAL, which in addition to a monetary reward gives them  voting power in Balancer's governance. This is again a point of centralization in the market which makes the platforms co-dependent, meaning a collapse in one platform can pull down the other. 



\section{Related Work}
\label{section:relatedWork}
Although governance protocols underline many prominent DeFi projects (such as Uniswap, Aave, and the others which were covered in this work), the literature dedicated to the subject is rather sparse.
Kiayias~\emph{et al.}~\cite{kiayias2022sok} present a SoK of governance protocols, which defines several properties of interest such as who is granted suffrage by each protocol, and the confidentiality of the voting process. They then perform a qualitative analysis of both the governance of blockchain consensus mechanism (e.g., the method by which improvements are suggested and adopted by cryptocurrencies such as Bitcoin and Ethereum), and the on-chain protocols used by DeFi protocols.
Barbereau~\emph{et al.}~\cite{barbereau2022defi} quantify the decentralization of the governance mechanisms of Uniswap, Maker, SushiSwap, Yearn Finance, Universal Market Access by applying various metrics to the governance token balances of the aforementioned mechanisms' users. Sun \emph{et al.}~\cite{sun2022decentralization} attempt to quantify the level of centralization in MakerDao's governance.
Fritsch \emph{et al.}~\cite{fritsch2022analyzing} map governance holding power and clusters that vote together for Uniswap and compound.
Barbereau \emph{et al.}~\cite{barbereau2022decentralised} analyze initial distribution of governance tokens and centrality in voting. 
Fan \emph{et al.}~\cite{fan2022towards} study how issuing governance tokens affects user's incentives to participate in a DeFi platform, mainly to provide liquidity.

\section{Conclusions} \label{section:conclusions}
This studies the design of popular governance mechanisms and analyzes points of centralization in protocols that are meant to be decentralized. It shows that participation rates are low, users tend to vote even less at times of high transaction fees, and voting is highly centralized. We identify specific examples of users who hold governance tokens for the duration of a single proposal which they created and voted for with a majority of the voting power. These proposals were retroactively discovered as harmful to the platform, and reverted at a later date. From all of these a picture is painted, implying that governance tokens are often not fulfilling the goal for which they are meant. Specifically, users of the platform do not use them to vote for proposals that create a more stable and safe platform overtime, but rather often use them to promote their own goals, which are sometimes harmful to the platform itself. These results align with ample research on referendums~\cite{feddersen2004rational}. The delegation model might be a remedy to some of the issues discussed in this paper. It however does not solve the incentives problem of users losing money by voting instead of investing their governance token. Exploring mechanisms that combine a delegation model with financial incentives is therefore a good direction for future work.

\newpage
\bibliographystyle{plain}
\bibliography{bibliography}


\end{document}

%% file: preamble.tex
\usepackage{mathtools}      
\usepackage{subfig}
\usepackage{lipsum}
\usepackage{enumitem}       
\usepackage{graphicx}       
\graphicspath{ {./images/} }
\usepackage{tabularx}       
\usepackage{makecell}       

\usepackage{amsthm}         
\usepackage{thmtools,thm-restate}   
\usepackage{hyperref} 
\usepackage[capitalise]{cleveref}       
\usepackage{subfiles} 

\newif\iflong   
\newif\ifmorecites
\newif\ifanonymous

\newif\ifcomments   
\usepackage[colorinlistoftodos,prependcaption,textsize=tiny,textwidth=\marginparwidth]{todonotes}

\newtheorem*{example*}{Example}

\crefname{definition}{Definition}{Definitions}
\crefname{theorem}{Theorem}{Theorems}
\crefname{claim}{Claim}{Claims}
\crefname{lemma}{lemma}{lemmas}
\crefname{corollary}{Corollary}{Corollarys}
\crefname{example}{Example}{Examples}
\crefname{remark}{Remark}{Remarks}

\crefname{code}{Code}{Code}